\documentclass{Pos-mod}

\def\beq{\begin{equation}}
\def\eeq{\end{equation}}

\title{Searches for Light Scalars, Pseudoscalars, and Gauge Bosons}

\ShortTitle{}

\author{\speaker{Abner Soffer} \\ 
        Tel Aviv University, Tel Aviv 69978, Israel\\
        E-mail: \email{asoffer@tau.ac.il}}

\abstract{ In the past few years there has been a great deal of
  theoretical and experimental activity related to the search for
  low-mass scalars, pseudoscalars, and vectors in various scenarios of
  physics beyond the standard model. I review the current status of
  this topic, focusing on results obtained since FPCP 2014.}

\FullConference{Flavor Physics \& CP Violation 2015,\\
		May 25-29, 2015\\
		Nagoya, Japan}

\begin{document}

\section{Introduction and motivation}
A variety of new-physics scenarios allow for new, weakly interacting
GeV-scale bosons. A class of scenarios involves a ``hidden sector'', a
loose term referring to a collection of theoretically related
particles, possibly with a rich phenomenology internal to that sector,
that are hidden from us due to the lack of a strong interaction with
the particles of the standard model (SM).
The gravitationally observed dark matter in the universe may be part of 
such a hidden sector, which is then called the dark sector. 
This term is often used even when one does not enforce a relation
to the observed dark matter, and I will do so here as well.
Of particular interest to particle physicists is the possible
existence of weak interactions -- so-called ``portals'' -- that allow
dark-sector particles to be produced in experiments and to decay via
detectable signatures.

In what follows, I briefly describe several scenarios in which
low-mass scalar, pseudoscalar, and vector bosons arise, and present
the results of recent searches for such particles.

\section{Searches for a dark photon}
\begin{figure}[!hbtp]
\begin{center}
\includegraphics[width=0.4\textwidth]{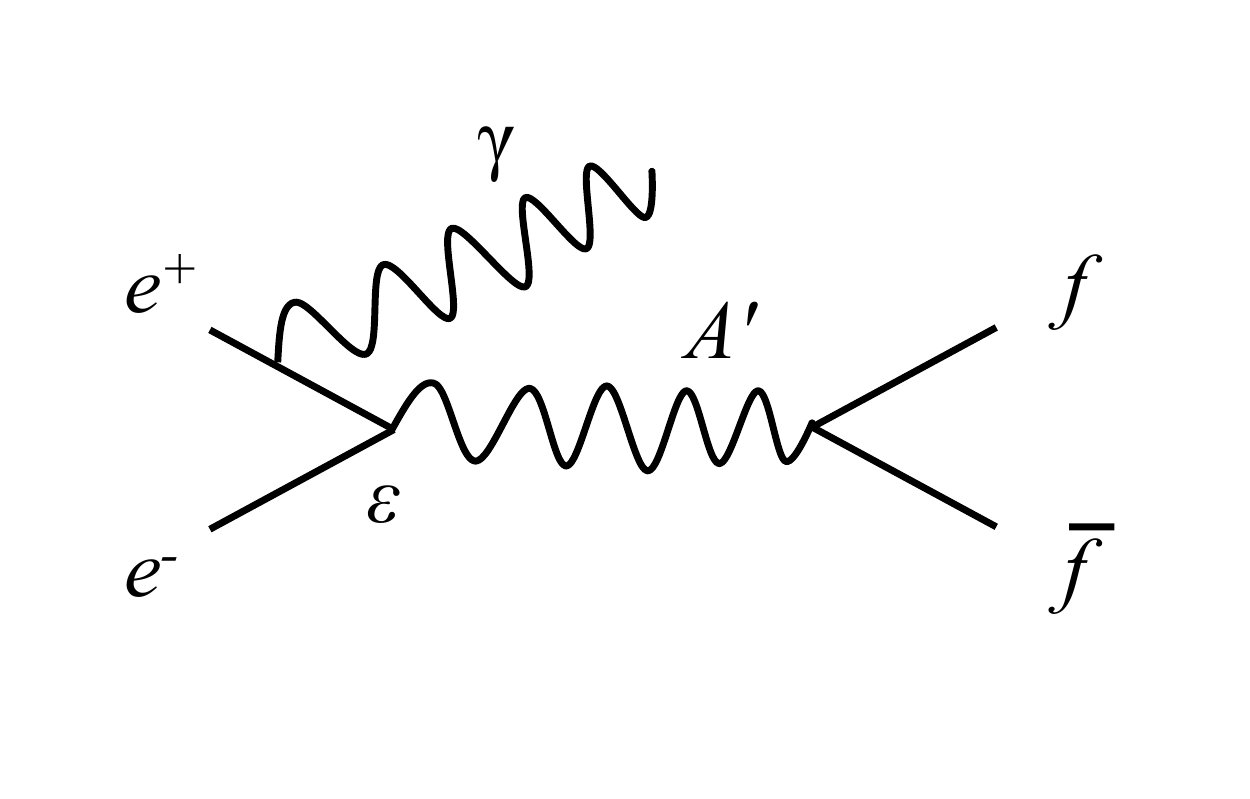}
\caption{\label{fig1} The production of an on-shell dark photon $A'$
  in an $e^+ e^-$ collision. The cross section is governed by the
  kinetic-mixing parameter $\epsilon$. The dark photon subsequently
  decays into two SM fermions via the same process.}
\end{center}
\end{figure}

A U(1)' gauge interaction in the dark sector gives rise to a dark
photon $A'$. The $A'$ may obtain mass via breaking of the symmetry,
distinguishing it from the SM photon. A so-called ``vector portal''
between the SM and the dark sector is provided by a kinetic mixing
term in the effective Lagrangian, 
${\cal L} \supset  -{1 \over 2} \epsilon F^{\mu\nu} F'_{\mu\nu}$,
where  $F'_{\mu\nu} = \partial_\mu A'_\nu - \partial_\nu A'_\mu$, 
$F^{\mu\nu}$ is the corresponding field
tensor for the SM U(1) interaction, and $\epsilon$ is the mixing
parameter between the two U(1) interactions. The Lagrangian gives rise
to processes shown in Fig.~\ref{fig1}, in which the dark photon is 
created on-shell in radiative electron-positron collisions.

Interest in dark photons was generated a few years ago, when the
PAMELA satellite experiment found that the ratio of positron and
electron energy spectra in cosmic rays was rising with energy for
energies above about 10~GeV~\cite{Adriani:2008zr}. This behavior, which
was later confirmed by Fermi~\cite{FermiLAT:2011ab} and
AMS-02~\cite{Aguilar:2013qda}, was interpreted as a possible result of
GeV-scale dark-photon production from the annihilation of TeV-scale
dark-matter fermion pairs~\cite{ArkaniHamed:2008qn}.
The observed spectrum was also shown to be consistent with the
expectation from secondary cosmic-ray production, without requiring
any new physics~\cite{Blum:2013zsa}.  Nonetheless, this observation
raised awareness for the possibility of light hidden-sector photons,
prompting a wealth of theoretical and experimental activity.

The latest dark-photon search was performed by the BES-III
collaboration, using the processes $e^+ e^- \to \gamma \ell^+ \ell^-$,
where $\ell$ is an electron or a muon~\cite{ref:bes}. In this so-called
untagged approach, observation of the initial-state-radiation (ISR)
photon was not required, in order to increase the efficiency and
accept events in which the photon was too forward to be detected. The
dilepton invariant-mass spectrum was fit with a 4th-order polynomial
for the background, plus a signal peak whose position was moved
throughout the spectrum. This treatment of the signal mass is
referred to as a signal ``scan''. No significant signal was found, and
limits were set on the value of $\epsilon$ as a function of $m_{A'}$.

\begin{figure}[!hbtp]
\begin{center}
\includegraphics[width=0.85\textwidth]{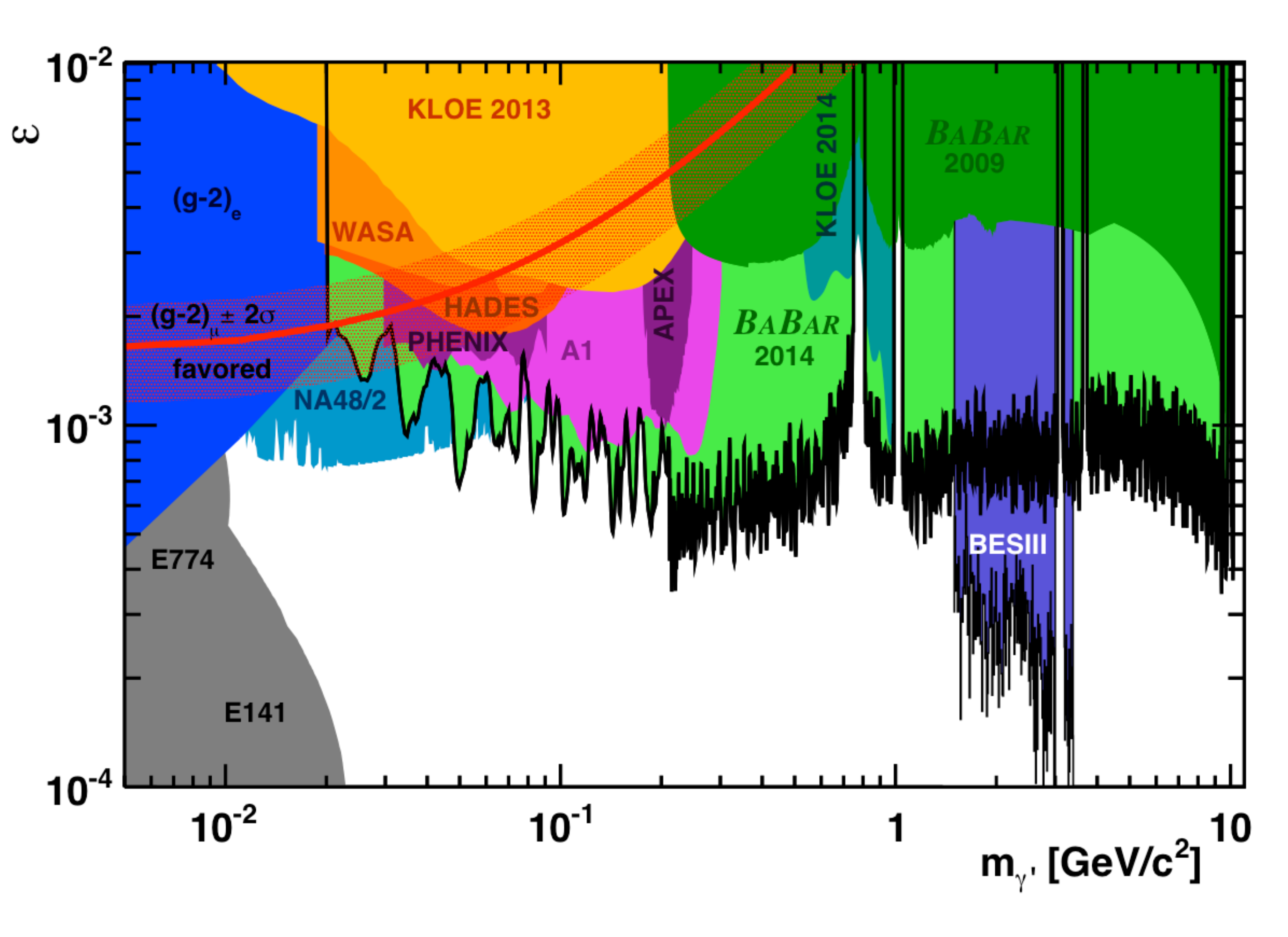}
\caption{\label{fig2} Upper limits on the mixing parameter $\epsilon$
  as a function of the dark-photon mass $m_{A'}$ from a variety of
  experiments.  See text for details (Figure: B. Echenard.)}
\end{center}
\end{figure}

The BES-III limits are shown in Fig.~\ref{fig2}. Also shown are all
previous results obtained from different experiments with a variety of
methods. These include $e^+ e^- \to \gamma \ell^+ \ell^-$ at
BABAR~\cite{Lees:2014xha} and at KLOE~\cite{Babusci:2014sta}, $\phi\to
\eta e^+ e^-$ at KLOE~\cite{Babusci:2012cr}, $\pi^0\to \gamma e^+ e^-$
at WASA-at-COSY~\cite{Adlarson:2013eza} and
NA48~\cite{Batley:2015lha}, $\pi^0\to \gamma e^+ e^-$ and $\eta\to
\gamma e^+ e^-$ at PHENIX~\cite{Adare:2014mgk}, and inclusive $e^+e^-$
spectra from proton-target interactions at
HADES~\cite{Agakishiev:2013fwl} and from electron-target interactions
at A1~\cite{Merkel:2014avp} and at APEX~\cite{Abrahamyan:2011gv}.
Also shown are constraints from measurements of the 
electron anomalous magnetic moment~\cite{Endo:2012hp}, as well as the 
region of $\epsilon, m_{A'}$ parameter space preferred by the 
discrepancy between the predicted and measured values of the muon 
anomalous magnetic moment~\cite{Pospelov:2008zw}. As shown in the figure, 
that preferred region is now fully excluded by other measurements.
The very low-$\epsilon$, low-$m_{A'}$ region is excluded by
reinterpretations of older beam-dump
experiments~\cite{Blumlein:2011mv,Andreas:2012mt,Blumlein:2013cua}.

It is worthwhile to consider some improvements to these studies in the
near future and beyond. First, I note that the BABAR
analysis~\cite{Lees:2014xha} was ``tagged'', i.e., used events in
which the ISR photon was observed. By contrast, the untagged approach
of the BES-III analysis~\cite{ref:bes} made it possible to utilize 
events in which the ISR photon is emitted at too
small a polar angle to be observed.  It is likely that BABAR can
tighten its current limits for regions of $m_{A'}$ by performing an
independent untagged analysis.

Belle can perform the same analysis as BABAR, with double the
integrated luminosity $L$. However, since the analysis is
background-dominated, and the signal yield is proportional to
$\epsilon^2$, the sensitivity to $\epsilon$ is proportional to
$L^{-1/4}$. Belle-II will have an integrated luminosity about 100
times that of BABAR, as well as better mass resolution due to the
larger drift chamber, and a more efficient $e^+ e^-$ trigger.  This
combination of factors will significantly increase the sensitivity
at better than the 4th-root of the ratio of luminosities~\cite{hearty}.

Additional dark-photon searches are planned by the Jefferson-Lab
experiments LIPSS, DarkLight, HPS, and APEX~\cite{Boyce:2012ym},
as well as VEPP-3 and A1~\cite{hearty}.

\section{Searches for a dark photon with dark Higgs}
\begin{figure}[!hbtp]
\begin{center}
\includegraphics[width=\textwidth]{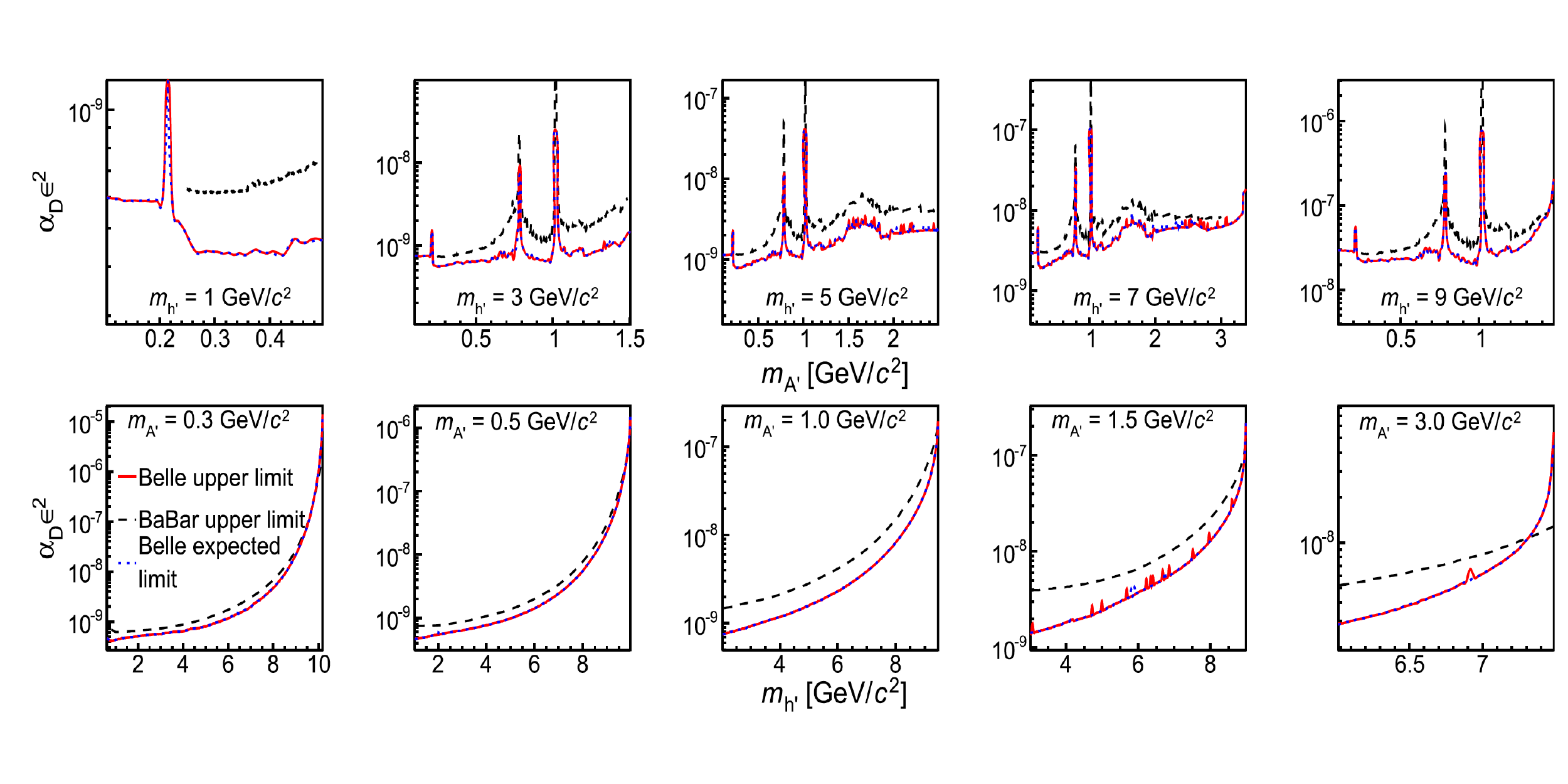}
\caption{\label{fig4} Upper limits on the product $\epsilon^2
  \alpha_D$ from the Belle search for production of three dark photons
  in $e^+ e^-$ collisions~\protect{\cite{TheBelle:2015mwa}} (solid red
  curves). Also shown are earlier results from a similar BABAR
  analysis~\protect{\cite{Lees:2012ra}} (dashed black curves).  }
\end{center}
\end{figure}

The next step up in model complexity includes also a light Higgs
scalar $h'$.  If its mass satisfies $m_{h'} > 2 m_{A'}$, the $h'$ can
decay into two dark photons. The Belle collaboration has recently
searched for this process in the process $e^+ e^- \to A'^*$, with
$A'^*\to A' h'$ and $h'\to A'A'$~\cite{TheBelle:2015mwa}. The three
final-state dark photons were searched for in their decays to pairs of
leptons or charged pions, or to two lepton pairs and an inclusive
final state $X$ whose mass was determined from the missing mass in the
event. Background was greatly suppressed by requiring the masses of
the three $A'$ candidates to be similar. The number of candidate
events observed was was consistent with the expected background, and
Belle set limits on the product of $\epsilon^2$ and the dark-sector
coupling constant $\alpha_D$. These limits are shown in
Fig.~\ref{fig4}.

\section{Searches for a light Higgs}
\begin{figure}[!hbtp]
\begin{center}
\includegraphics[width=0.4\textwidth]{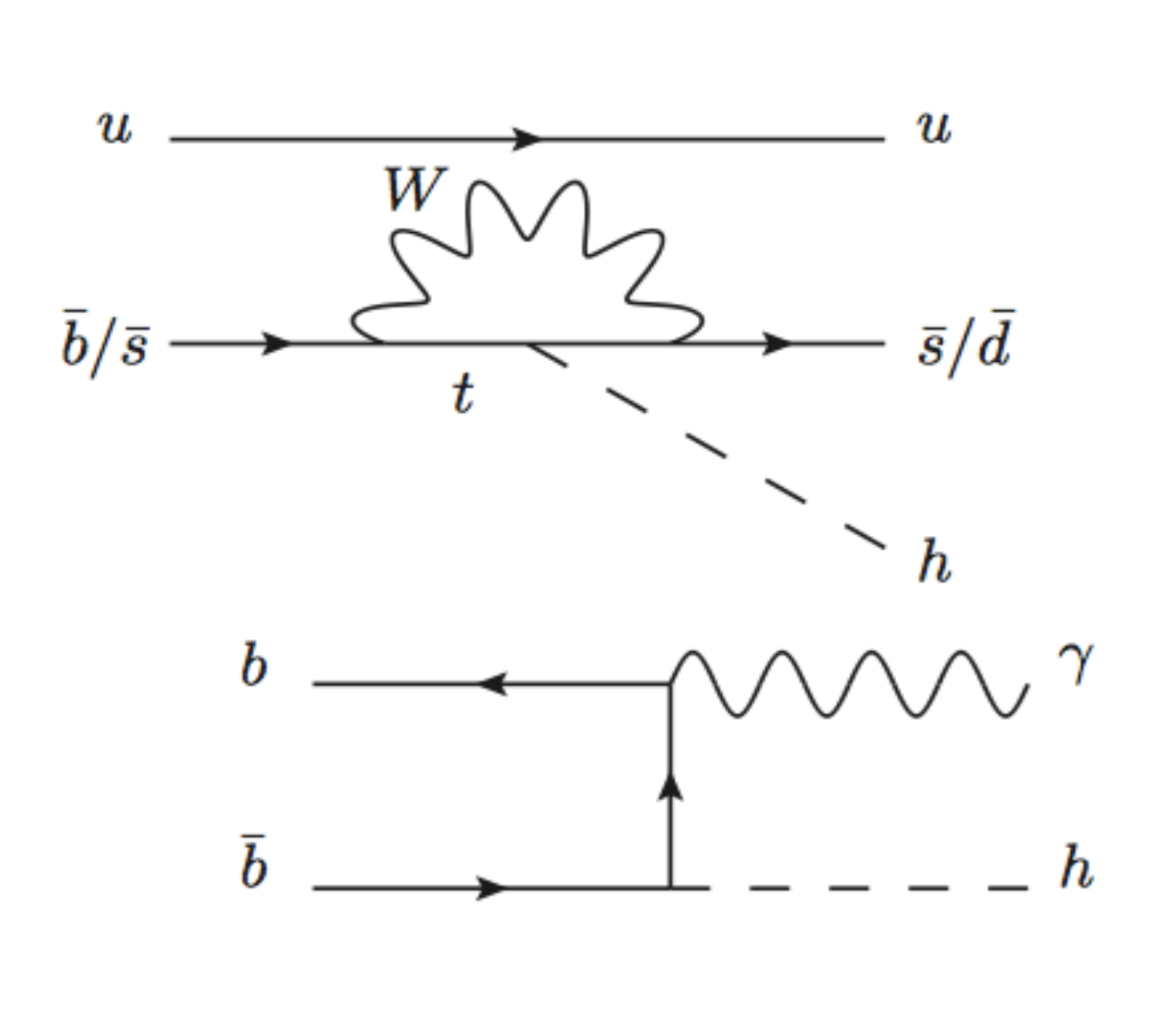}
\caption{\label{fig5} Diagrams for the production of a light scalar $h$
via its large coupling to (top) the top quark in penguin $B$-meson decays
or (bottom) to the bottom quark in radiative $\Upsilon$ decays.
}
\end{center}
\end{figure}

A light Higgs also comes up in scenarios that do not involve a dark
photon.  One example is the next-to-minimal supersymmetric standard
model (NMSSM)~\cite{Dermisek:2006py}, which contains a light CP-odd
Higgs.  Another is a case where a light scalar mixes with
the SM Higgs~\cite{Schmidt-Hoberg:2013hba, Clarke:2013aya}, called
``Higgs portal'' in the case of a dark-sector Higgs. In either case,
the light scalar may be produced in decays of a $B$ or $\Upsilon$
meson, as shown in Fig.~\ref{fig5}, taking advantage of the large
Higgs couplings to the heavy top and bottom quarks.  

Depending on the scenario, different symbols are used for the light
Higgs. In what follows, I use the generic symbol $h$.

The CLEO collaboration was the first to perform a search for a light
Higgs, using radiative decays of the $\Upsilon(1S)$ and the light-Higgs decay
channels $h\to \mu^+ \mu^-$ and $h \to \tau^+
\tau^-$~\cite{Love:2008aa}.
BABAR has used decays of both the 
$\Upsilon(3S)$~\cite{Aubert:2009cp, Aubert:2009cka, Lees:2011wb, Aubert:2008as} and the
$\Upsilon(1S)$~\cite{Lees:2012iw, Lees:2012te, Lees:2013vuj, delAmoSanchez:2010ac}, searching for light-Higgs decays into $\mu^+ \mu^-$,
$\tau^+ \tau^-$, hadrons, and invisible particles. 
BES-III has done the same with $J/\psi$ decays in the mode
$h\to\mu^+\mu^-$~\cite{Ablikim:2011es}.
CMS has performed both an inclusive search in the $h\to \mu^+\mu^-$
channel~\cite{Chatrchyan:2012am} and an exclusive search in decays of
the 125-GeV Higgs, $H\to hh\to
\mu^+\mu^-\mu^+\mu^-$~\cite{Chatrchyan:2012cg}.
In a very recent analysis, ATLAS has searched for $H\to hh\to
\mu^+\mu^-\tau^+\tau^-$~\cite{Aad:2015oqa}. Please see Peter Onyisi's
contribution to this conference for details.
The combination of all these searches places tight constraints on the
NMSSM scenario.

The branching fractions for decays of the $h$ to the different final
states depend on its mass and on the parameters of the
model. Therefore, it is important to search in all possible
channels. Until recently, the only light-Higgs decays that had not yet
been explored were decays into heavy quarks. This has now been
addressed by BABAR for the $2m_D < m_h<2 m_B$ case, in a new search for $h\to
c\bar c$, where the $h$ is produced in the decay $\Upsilon(1S) \to
\gamma h$~\cite{Lees:2015jwa}.

\begin{figure}[!hbtp]
\begin{center}
\includegraphics[width=0.65\textwidth]{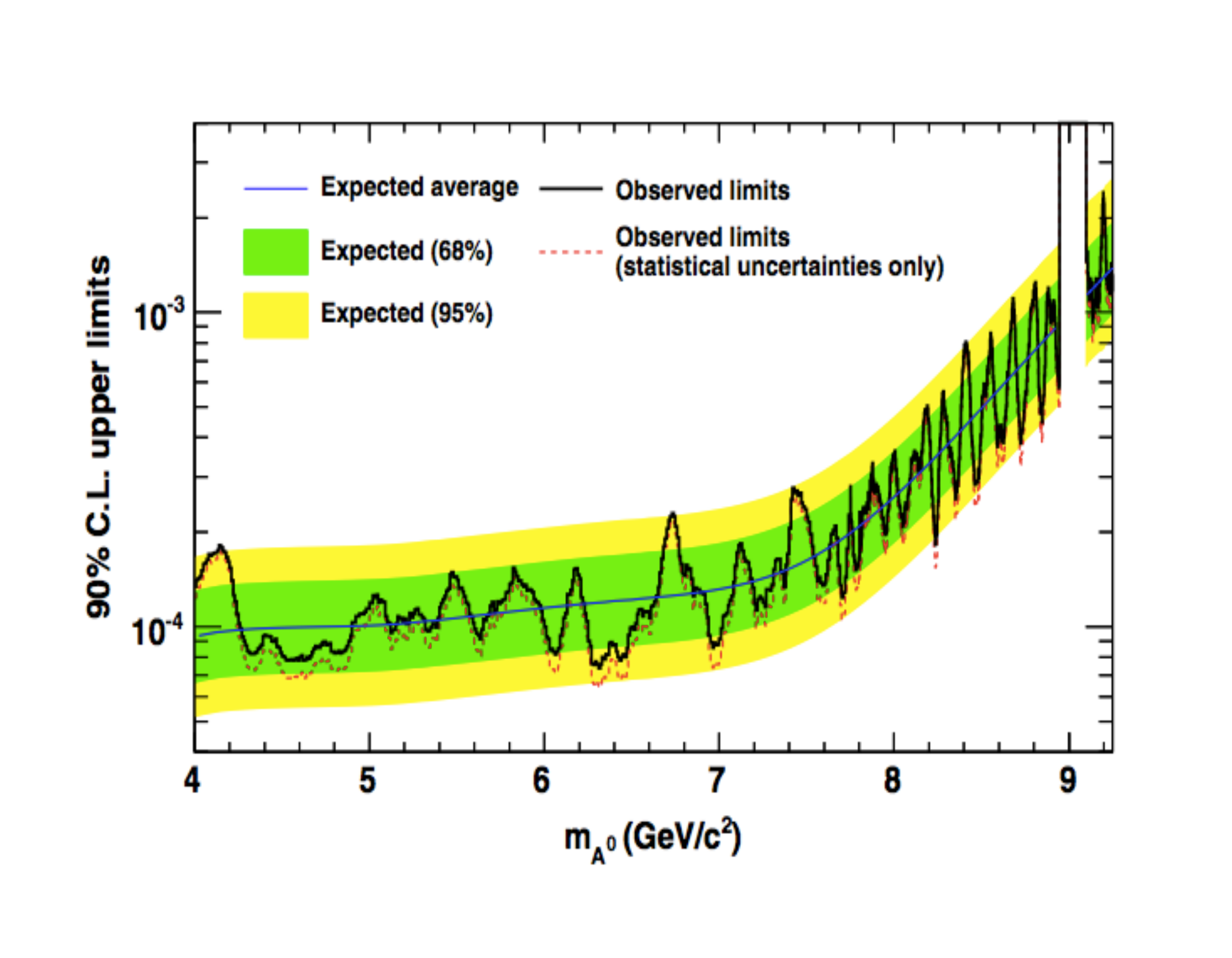}
\caption{\label{fig6} Upper limits at the 90\% confidence level on the 
the product of
branching fractions ${\cal B}(\Upsilon(1S)\to \gamma h)\times {\cal
  B}(h\to c\bar c)$ from BABAR~\protect{\cite{Lees:2015jwa}}.}
\end{center}
\end{figure}

The analysis used a sample of $121\times 10^6$ $e^+e^-\to
\Upsilon(2S)$ events, with the decay $\Upsilon(2S)\to \pi^+ \pi^-
\Upsilon(1S)$ used to obtain a high-purity sample of $\Upsilon(1S)$
mesons. The square of the mass recoiling against the pions,
$(p_{e^+e^-} - p_{\pi^+\pi^-})^2$, where $p_{e^+e^-}$ is the
4-momentum of the incoming beam particles and $p_{\pi^+\pi^-}$ is that
of the reconstructed pion pair, was required to be consistent with the
known mass of the $\Upsilon(1S)$. A photon and a charmed meson were then
reconstructed, thus tagging the decay  $h\to c\bar c$. The 
mass of the $h$ candidate was calculated from 4-momentum conservation,
$m_h^2=(p_{e^+e^-} - p_{\pi^+\pi^-} - p_\gamma)^2$, where $p_\gamma$
is the 4-momentum of the photon. 
The resulting spectrum of $h$ candidates was then fit with a
polynomial background, and a signal-peak scan was performed. No
significant signal was observed, and limits were set on the product of
branching fractions ${\cal B}(\Upsilon(1S)\to \gamma h)\times {\cal
  B}(h\to c\bar c)$. These limits are shown in Fig.~\ref{fig6}.

\section{Long-lived particles}
New hidden-sector particles may be weakly interacting enough to be
long-lived, so that they can be identified as displaced vertices.
In fact, several low-$\epsilon$, low-$m_{A'}$ limits in the dark-photon
parameter space are based on searches for long-lived particles in beam-dump
experiments~\cite{Blumlein:2011mv,Andreas:2012mt,Blumlein:2013cua}.

At colliders, which have inherently lower luminosities than beam-dump
experiments, observing such a signature requires that the new particle
be produced relatively strongly and only decay very weakly. This is
achieved if there exist two different portals with different
strengths.  Alternatively, the Higgs portal offers such a mechanism,
since scalar-fermion couplings are proportional to the fermion
mass. Thus, a light Higgs may be produced via a large Yukawa coupling
to heavy fermions, as in Fig.~\ref{fig5}, yet decay with a small
Yukawa coupling if its small mass allows decays only to muons or
electrons.
An interesting model of this type is that of a GeV-scale inflaton $X$ that
mixes with the SM Higgs~\cite{Bezrukov:2013fca} via the quartic term
${\cal L}\supset -\lambda \left(H^\dagger H - {\alpha \over \lambda} X^2\right)^2$. 
Depending on the
values of the model parameters, the average inflaton may either decay
promptly, form a displaced vector in the detector, or decay outside
the detector.

\begin{figure}[!htbp]
\begin{center}
\begin{tabular}{cc}
\hspace{-0.8cm} 
\includegraphics[width=0.6\textwidth]{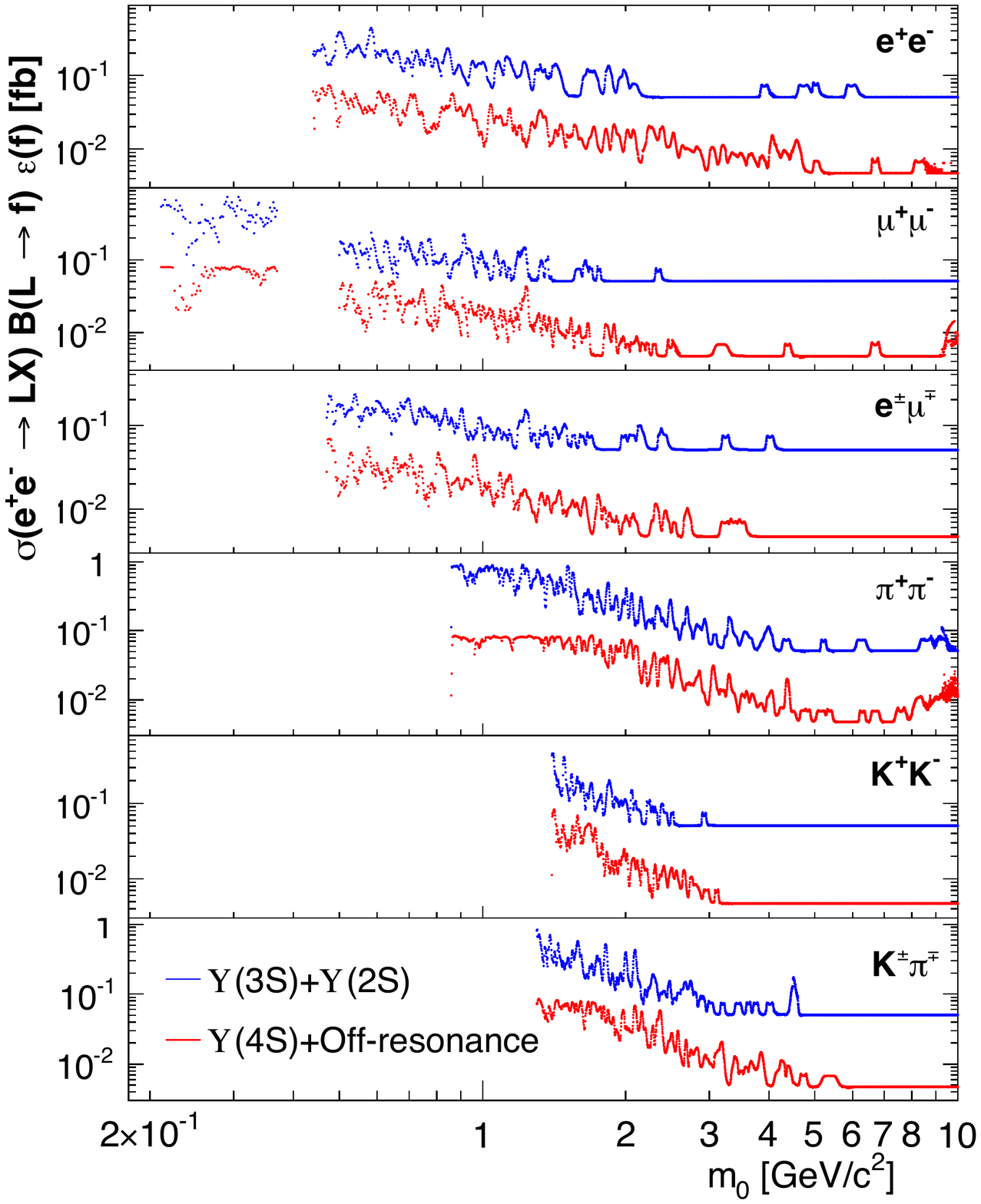} &
\hspace{-2cm} 
\includegraphics[width=0.6\textwidth]{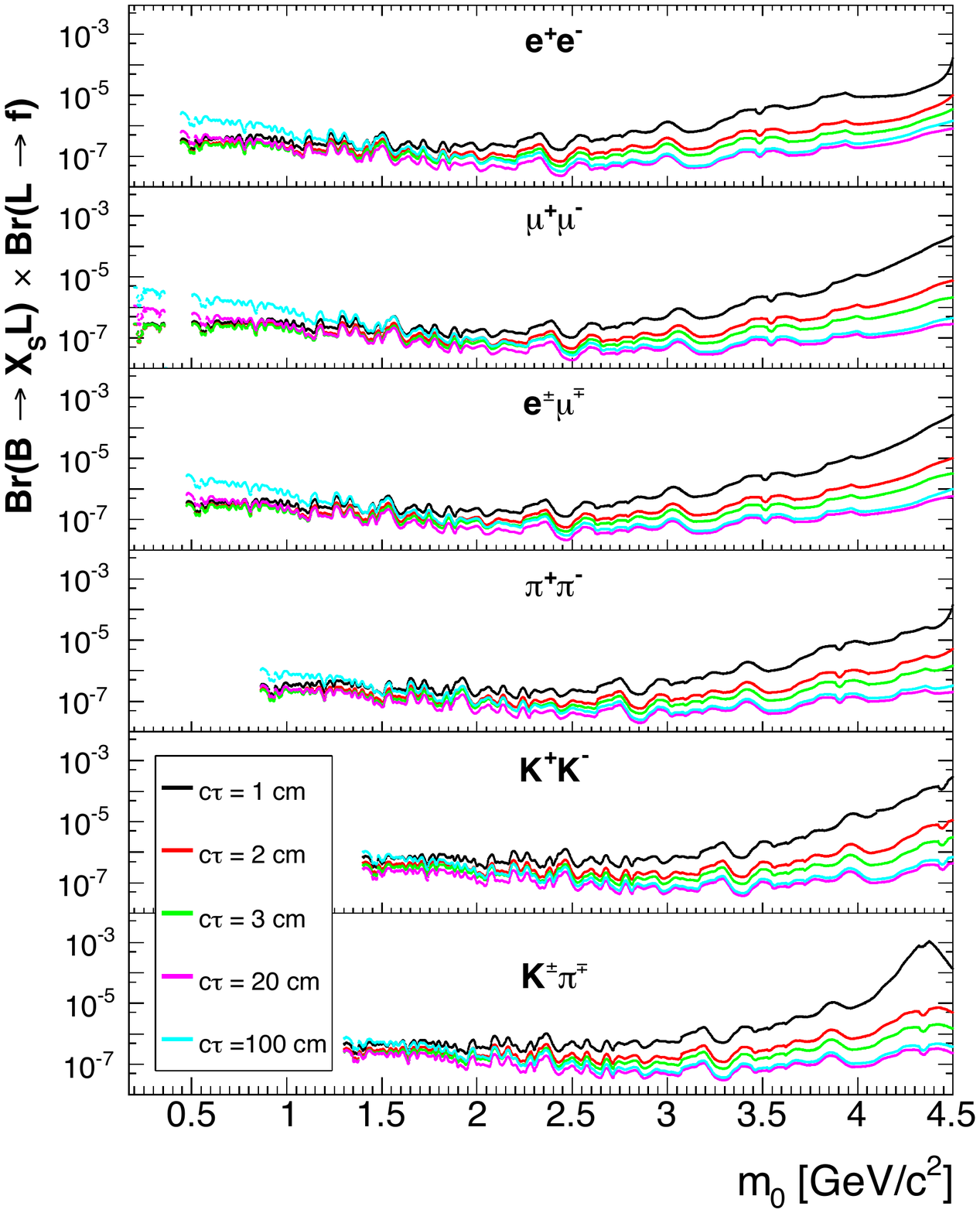} 
\end{tabular}
\caption{\label{fig7} Results of the BABAR search for a long-lived
  particle $L$~\protect{\cite{Lees:2015rxq}}. (Left)
  ``Model-independent'' limits on the product of the $L$ production
  cross section, decay branching fraction into each of the 6 final
  states shown, and efficiency for reconstruction of that final state.
  These limits are intended for recasting for specific models,
  calculating the total efficiency using efficiency tables provided in
  Ref.~\protect{\cite{Lees:2015rxq}}. (Right) ``Model-dependent''
  limits in the Higgs-portal scenario of $L$ production in penguin $B$
  decays.  }
\end{center}
\end{figure}

BABAR has recently searched for inclusive production of a generic
long-lived particle $L$ that decays into two leptons or two charged
hadrons, forming a displaced vertex~\cite{Lees:2015rxq}. Under the
hypothesis that the $L$ is fully reconstructed, the search consisted
in scanning for a peak in the displaced-vertex mass spectrum. No
significant signal was found, and the results were reported as two
types of limits.
In the ``model-independent'' presentation, upper limits were presented
for the product $\sigma(e^+ e^-\to L) \, {\cal B}(L\to f) \,
\epsilon(f)$ of the $L$ production cross section, the branching
fraction for its decay into each final state $f$ under study, and the
reconstruction efficiency for $f$.  Efficiency tables in terms of the
$L$ mass, its lifetime, and transverse momentum were provided, to enable
recasting of the results to any model.
In the ``model-dependent'' presentation, the efficiency was calculated
for $B\to L X_s$ decays, corresponding to the top diagram in
Fig.~\ref{fig5}, where $X_s$ is a hadronic state with strangeness.
Limits on ${\cal B}(B\to LX_s)\, {\cal B}(L\to f)$ were then presented.
The two types of limits are shown in Fig.~\ref{fig7}.

Another possibility for production of long-lived particles is that
they are created in decays of heavy, more strongly interacting
particles. An example is the model of Ref.~\cite{Falkowski:2010cm},
where the 125~GeV Higgs decays into two hidden-sector fermions, which
decay to the lightest hidden fermion by emitting hidden-sector
photons. Being the lightest hidden-sector state, the hidden photon can
decay only via kinetic missing into lepton pairs, and its lifetime is
long.  The resulting signature is that of displaced lepton jets.

\begin{figure}[!hbtp]
\begin{center}
\includegraphics[width=0.9\textwidth]{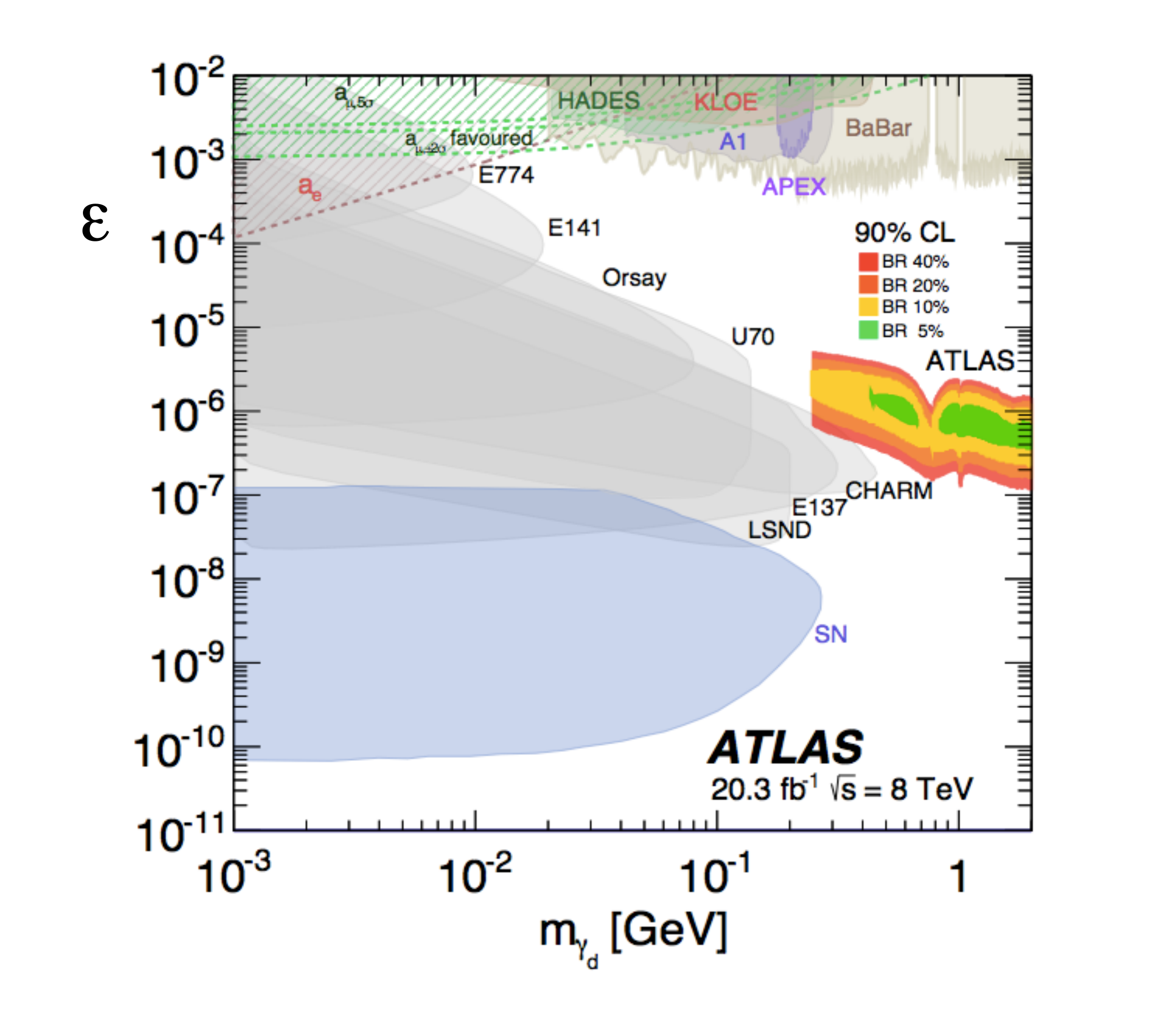} 
\caption{\label{fig8} The $\epsilon$ vs. $m_{A'}$ parameter region
  excluded by ATLAS~\protect{\cite{Aad:2014yea}} within the context of
  the model of Ref.~\protect{\cite{Falkowski:2010cm}}. Also seen are 
constraints from other results. See text for details.
}
\end{center}
\end{figure}

ATLAS has searched for this signature, requiring two lepton jets that
are separated by a large azimuthal angle and are well isolated from
additional high-transverse-momentum particles in the
event~\cite{Aad:2014yea}.  The background estimation was based on
sidebands of the angle and the isolation variable. Results of this 
search are presented in Fig.~\ref{fig8} as an excluded region in the
space of $\epsilon$ vs. $m_{A'}$, for given hypotheses regarding the
other model parameters, such as the masses of the hidden-sector
fermions. 
Also shown are limits from other experiments that appear in
Fig.~\ref{fig2}, as well as constraints from additional
fixed-target experiments (Orsay~\cite{Davier:1989wz},
U70~\cite{Blumlein:2013cua},
CHARM~\cite{Gninenko:2012eq},
LSND~\cite{Essig:2010gu},
E137~\cite{Blumlein:2011mv,Andreas:2012mt,Blumlein:2013cua})
and constraints derived from the supernova cooling 
rate~\cite{Dent:2012mx, Dreiner:2013mua}.

\section{Search for a $\pi^0$ impostor}
In 2009, BABAR found~\cite{Aubert:2009mc} that the $\gamma\gamma^*\to
\pi^0$ transition form factor exceeded the expected high-$Q^2$
asymptotic value of $185~{\rm MeV}/Q^2$~\cite{Lepage:1979zb,
  Lepage:1980fj}.  Later results from Belle~\cite{Uehara:2012ag} were
consistent with both the QCD expectation and the BABAR excess.  The
possible discrepancy was given a new-physics explanation, in the form
of a ``$\pi^0$ impostor'' $\phi$ that couples only to $\tau$
leptons~\cite{McKeen:2011aa}. Couplings to other heavy fermions are in
principle also possible, but are excluded by existing measurements.
In order to explain the combined BABAR+Belle deviation from the
theoretical expectation, the cross section for $e^+ e^- \to \tau^+
\tau^- \phi$ has to be in the 95\% confidence level interval
$[50, 140]$~pb for a scalar $\phi$,
$[2.5, 5.1]$~pb for a pseudoscalar,
and $[0.15, 0.59]$~pb for a ``hard-core pion'', which is a
pseudoscalar state that also couples to light quarks and hence mixes
with the SM $\pi^0$.

\begin{figure}[!hbtp]
\begin{center}
\includegraphics[width=0.7\textwidth]{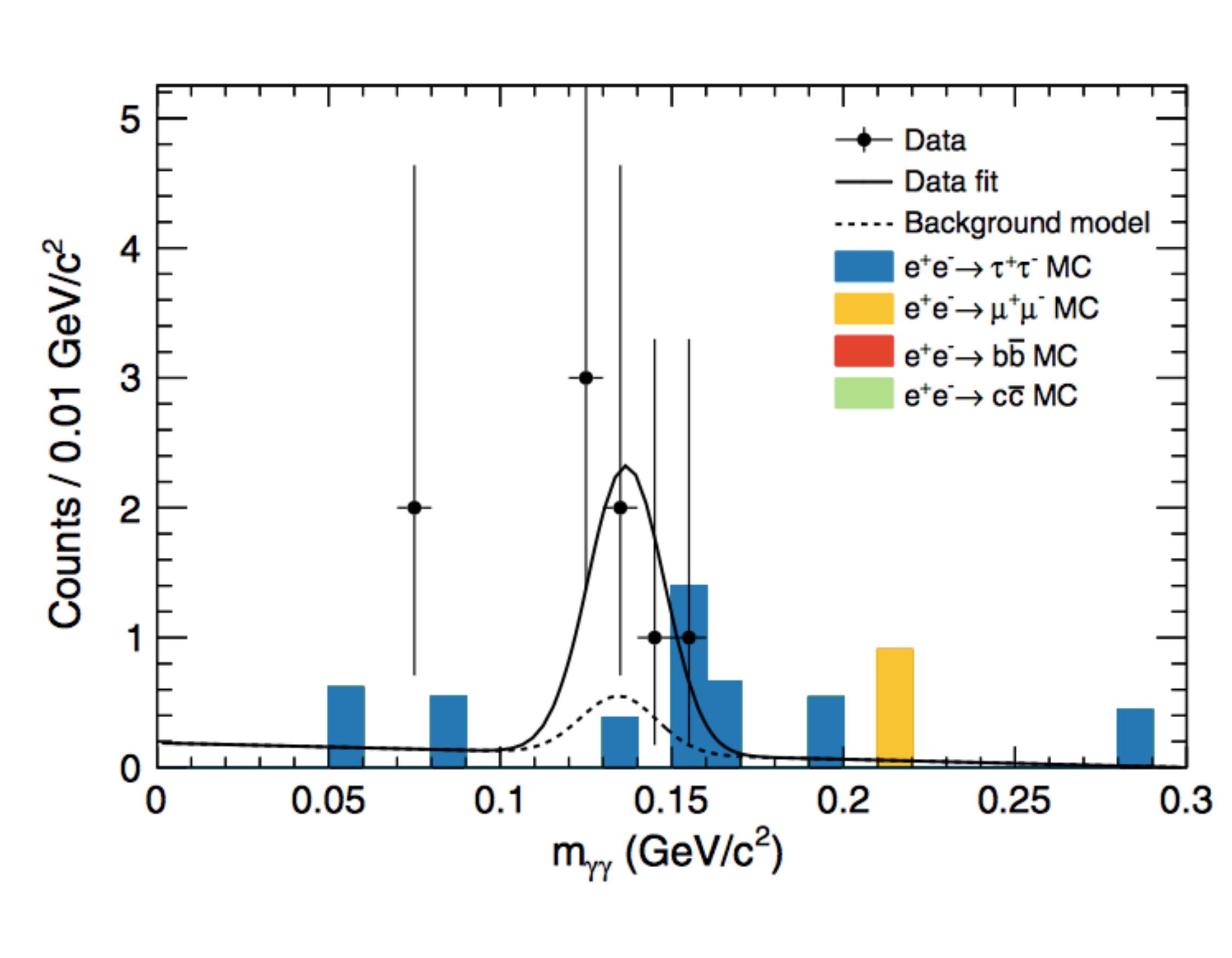} 
\caption{\label{fig9} The $\gamma\gamma$ spectrum (data points) seen
  by BABAR in the search for a $\pi^0$ impostor radiated from a $\tau$
  lepton~\protect\cite{BaBar:2014jfk}.  The solid curve shows
the best fit to signal plus background, with the dashed curve showing
the total background.}
\end{center}
\end{figure}

To test such a possibility, BABAR has searched for the process $e^+
e^- \to \tau^+ \tau^- \phi$~\cite{BaBar:2014jfk}. The $\tau^+\tau^-$
pair was identified in the final state $\mu^\pm e^\mp$ plus unobserved
neutrinos, and the $\phi$ was reconstructed in the decay $\phi\to
\gamma\gamma$.  The $\gamma\gamma$ mass spectrum was fit with a
polynomial background, and a signal-peak scan was performed. The
results of the fit with the largest signal yield are shown in
Fig.~\ref{fig9}. The search yielded an upper limits on the cross
section of $\sigma < 73$~fb for the pseudoscalar cases case and
$\sigma<370$~fb for a scalar impostor, thus ruling out the models of
Ref.~\cite{McKeen:2011aa}.

\section{Summary}
The results covered in this talk, all published in the past year, are
a testament to the interest in the physics of new GeV-scale particles.
Searches for such particles take place at a variety of facilities,
including  fixed-target experiments, the $B$ factories, and LHC,
providing sensitivity to a range of physics scenarios. With LHC
continuing to take data and new facilities coming online in the next
few years, this will continue to be an active area of research,
both theoretically and experimentally.

\section{acknowledgments}
I thank Hai-Bo Li, Youngjoon Kwon, and Bertrand Echenard for pointing
out relevant results and explaining the details of several analysis.
I also thank the organizers of FPCP-2015 for a well organized
conference, and the participants for interesting talks and useful
discussions.

\end{document}